\documentclass[prb,twocolumn,showpacs,preprintnumbers,amsmath,amssymb]{revtex4}

\usepackage{graphicx}
\usepackage{dcolumn}
\usepackage{bm}

\preprint{submitted to the Physical Review B}

\begin{document}

\title{Interface properties of the NiMnSb/InP and NiMnSb/GaAs contacts}

\author{Iosif Galanakis}\email{I.Galanakis@fz-juelich.de}
\affiliation{Institut f\"ur Festk\"orperforschung,
Forschungszentrum J\"ulich, D-52425 J\"ulich, Germany}
\author{Marjana Le\v{z}ai\'{c}}
\affiliation{Institut f\"ur Festk\"orperforschung,
Forschungszentrum J\"ulich, D-52425 J\"ulich, Germany}
\author{Gustav Bihlmayer}
\affiliation{Institut f\"ur Festk\"orperforschung,
Forschungszentrum J\"ulich, D-52425 J\"ulich, Germany}
\author{Stefan Bl\"ugel}
\affiliation{Institut f\"ur Festk\"orperforschung,
Forschungszentrum J\"ulich, D-52425 J\"ulich, Germany}

\begin{abstract}
  We study the electronic and magnetic properties of the interfaces
  between the half-metallic Heusler alloy NiMnSb and the binary
  semiconductors InP and GaAs using two different state-of-the-art
  full-potential \textit{ab-initio} electronic structure methods.
  Although in the case of most NiMnSb/InP(001) contacts the
  half-metallicity is lost,
  it is possible to keep a high degree of spin-polarization when the
  interface is made up by Ni and P layers. In the case of the GaAs
  semiconductor the larger hybridization between the Ni-$d$ and As-$p$
  orbitals with respect to the hybridization between the Ni-$d$ and
  P-$p$ orbitals destroys this polarization.  The (111) interfaces
  present strong interface states but also in this case there are few
  interfaces presenting a high spin-polarization at the Fermi level
  which can reach values up to 74\%.
\end{abstract}



\maketitle

\section{Introduction
\label{sec1} }

The spin-injection from a metal into a semiconductor remains one of
the main challenges in the field of
magnetoelectonics.\cite{Zutic2004,Olaf1,Olaf2} The use of
half-metallic ferromagnets as electrodes was proposed to maximize the
efficiency of spintronic devices.  These compounds are ferromagnetic
metals with a band gap at the Fermi level ($E_F$) for the
minority spin band leading to 100\% spin-polarization at $E_F$. But
even in this case, interface states at the contact between the
half-metal and the semiconductor can destroy the half-metallicity.
Due to their orthogonality to all bulk states incident to the interface,
in the ballistic limit these states should not affect the transport
properties, but it is their interaction with other defect states which
makes them conducting.

The first material predicted to be a half-metal is the Heusler alloy
NiMnSb.\cite{groot} There exist several ab-initio calculations
on NiMnSb reproducing the initial results of de Groot and
collaborators,\cite{calc1,calc2,calc3,calc4,calc5} and Galanakis
\textit{et al.} showed that the gap arises from the hybridization
between the $d$ orbitals of the Ni and Mn atoms.\cite{iosifHalf}
Experiments seem to well establish its half-metallicity in the case of
single crystals,\cite{Kir,Hans} but in films the
half-metallicity is lost.\cite{Molenkamp1,Molenkamp2,film1,film2,film3,film4,film5,film6,film7}
Theoretical calculations for the interfaces of these materials with
the semiconductors are few and all results agree that in general the
half-metallicity is lost at the interface between the Heusler alloy and the
semiconductor.\cite{groot2,Debern,PicozziInter1,PicozziInter2,iosifCo2CrAl}
Wijs and de Groot have argued than in the case of the NiMnSb/CdS (111)
contacts the Sb/S interface keeps the half-metallicity (or at least
shows a very high degree of spin-polarization) of the bulk
NiMnSb.\cite{groot2} Thus, even if half-metallicity is lost, it is
possible that a high degree of spin-polarization stays at the
interface and these structures remain attractive for realistic
applications.

We should also mention that even in the absence of the interface
states true half-metallicity can not really exist due to minority states
induced by the spin-orbit coupling which couples the two spin-bands.
But as it was shown for these systems in Refs. \onlinecite{Phivos1} and
\onlinecite{Phivos2} this phenomenon is very weak and instead of a
gap in the minority channel there is a region of still almost 100\%
spin-polarization. It was also found that the  orbital moments
are very small  in these compounds.\cite{iosifOrbit} Thus, spin-orbit
coupling can be assumed to  be negligible with respect to the interface
states.

In this communication we study the (001) interfaces of the
half-metallic NiMnSb Heusler alloy with InP and GaAs and the (111)
interface between the NiMnSb and InP compounds.  We take into account
all possible contacts and show that there are cases where a high
degree of spin-polarization remains at the interface.  In Section
\ref{sec2} we discuss the structure of the interfaces and the
computational details and in Section \ref{sec3} we present and analyze
our results for the (001) interfaces. In Section \ref{sec4} we discuss
the (111) interfaces and finally in Section \ref{sec5} we summarize
and conclude.

\section{Computational method and structure
\label{sec2} }

In the calculations we used two different full-potential methods.
Firstly we employed the full-potential version of the screened
Korringa-Kohn-Rostoker (FSKKR) Green's function
method\cite{zeller95,Pap02} in conjunction with the local spin-density
approximation (LDA)\cite{vosko} to the density functional
theory\cite{Kohn1,Kohn2} to study the (001) interfaces between
NiMnSb and the InP and GaAs semiconductors. The FSKKR method
scales linearly with the number of atoms and, therefore, allows to
study also very thick slabs of these materials. But it cannot
give exactly the Fermi level of semiconductors due to problems arising
from the $\ell_{max}$ cut-off in this method\cite{Zeller2004} and, thus,
we employed also the full-potential linearized augmented plane-wave
method (FLAPW)\cite{Wimmer,Weinert} in the {\tt FLEUR}
implementation\cite{FLEUR} to calculate the NiMnSb/InP(001)
interfaces in order to compute the band offset. Finally, the FLAPW method
was also employed in the case of the NiMnSb/InP(111) interfaces.

NiMnSb crystallizes in the $C1_b$ structure, which consists of four
interpenetrating fcc sublattices. The unit cell is that of a fcc
lattice with four atoms as basis at A=$(0\:0\:0)$,
B=$({1\over4}\:{1\over4}\:{1\over4})$, C=
$({1\over2}\:{1\over2}\:{1\over2})$ and
D=$({3\over4}\:{3\over4}\:{3\over4})$ in Wyckoff
coordinates.\cite{iosifHalf} In the case of NiMnSb the A site is
occupied by Ni, the B site by Mn and the D site by Sb, while the C
site is unoccupied.  The $C1_b$ structure is similar to the $L2_1$
structure adopted by the full Heusler alloys, like Ni$_2$MnSb where
also the C site is occupied by a Ni atom.\cite{iosifFull} The
zincblende structure, adopted by a large number of semiconductors
like GaAs and InP, can also be considered as consisting of four fcc
sublattices. In the
case of GaAs the A site is occupied by a Ga atom, the B site by an As
atom, while the C and D sites are empty. Depending on the electronic
structure method used to perform the calculations one either uses
empty spheres or empty polyhedra to account for the vacant sites (as
it is done in the FSKKR) or the vacant sites just make part of the
interstitial region (as in FLAPW). Within 1\% accuracy NiMnSb (5.91\AA )
has the same experimental lattice constant as InP (5.87\AA)
and epitaxial growth of NiMnSb on top of InP has been already
achieved experimentally by molecular beam
epitaxy.\cite{Molenkamp1,Molenkamp2} On the other hand, the lattice
constant of GaAs (5.65\AA ) is almost 4\% smaller.  The dominant
effect at the interface is the expansion or the contraction of the
lattice of the half metal along the growth axis to account for the
in-plane change of its lattice
parameter.\cite{Debern,PicozziInter1,PicozziInter2} Since in the case
of the NiMnSb/InP interface both compounds have similar lattice
parameters, in the calculations perfect epitaxy can be assumed.

\begin{figure}
  \begin{center}
    \includegraphics[scale=0.4]{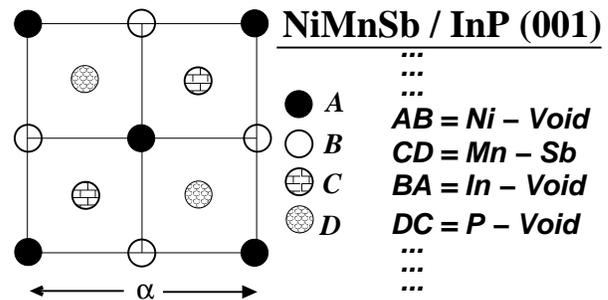}
  \end{center}
\caption{\label{fig1}
  Schematic representation of the (001) interface between NiMnSb and
  InP.  There are several different combinations at the interface
  which can be either Ni/In, Ni/P, MnSb/In (shown in the figure) or
  MnSb/P.}
\end{figure}

Within the FSKKR the space is divided into non-overlapping
Wigner-Seitz polyhedra and thus empty ones are needed to describe
accurately the vacant sites (similarly to the use of empty spheres in
the early electronic structure methods).  To simulate the (001)
interface within the FSKKR calculations we used a multilayer consisting
of 15 layers of the half-metal and 9 semiconductor layers.  This
thickness is enough so that the layers in the middle of both the
half-metallic part and the semiconducting one exhibit bulk properties.
There are several combinations at the interface, e.g.\ at the
NiMnSb/InP contact the interface can be either a Ni/In one, Ni/P,
MnSb/In or MnSb/P (see Fig.~\ref{fig1}).  We will keep this definition
through out the paper to denote different interfaces.  We should also
mention that since the multilayer contains 15 half-metal and 9
semiconductor layers, there are two equivalent surfaces at both sides of
the half-metallic spacer.  Finally, for our FSKKR calculations we used
a 20$\times$20$\times$4 grid in the \textbf{k}-space and we took into
account wavefunctions up to $\ell_{max}$=3 and thus the potential and
the charge density were expanded on lattice harmonics up to
$\ell_{max}$=6. All FSKKR calculations have been performed at the
experimental lattice constant of NiMnSb (5.91\AA ).

In the FLAPW method the space is divided into non-overlapping
muffin-tin spheres around each atom and an interstitial region,
that is described in terms of plane-waves. To perform the
calculations for the (001) interfaces we employed a repeated slab
made up of 8 layers of NiMnSb and 8 layers of the semiconductor.
Thus if the one contact is Ni/P the other one is MnSb/In. As will
be shown in subsection \ref{sec3-3}, the smaller number of layers
(as compared to the FSKKR calculations) does not influence the
properties near the Fermi level. For the (111) interfaces the
supercells consisted of 16 layers of NiMnSb and 12 layers of InP.
The FLAPW calculations were performed using density functional
theory in the generalized gradient approximation (GGA) as given by
Perdew et al.\cite{PBE} For the calculations, a planewave cutoff
$K_{\rm max}$ of 3.45~a.u.$^{-1}$ was used. Lattice harmonics with
angular momentum $l\leq 8$ were used to expand the charge density
and the wavefunctions within the muffin-tin spheres, having a
radius of 2.4~a.u. for Sb and 2.34~a.u. for all the other atoms.
The Brillouin-zone (BZ) was sampled with 128 special
\textbf{k}-points in the irreducible wedge (1/8 of the whole BZ)
for (001) interfaces, and 90 \textbf{k}-points in the irreducible
wedge (1/12 of the whole BZ) for the (111) interfaces. All FLAPW
calculations were performed at the experimental lattice constant
of InP (5.87~\AA).

\section{N\lowercase{i}M\lowercase{n}S\lowercase{b}/I\lowercase{n}P
  and N\lowercase{i}M\lowercase{n}S\lowercase{b}/G\lowercase{a}A\lowercase{s}
  (001) interfaces
\label{sec3} }

Compared to simple surfaces, interfaces are more complex systems
due the hybridization between the orbitals of the atoms of the
metallic alloy and the semiconductor at the interface. Thus, results
obtained for the surfaces (as the ones in Refs.~\onlinecite{iosifSurf}
and \onlinecite{iosif111}) cannot be easily generalized for interfaces
since for different semiconductors different phenomena can occur.  In
both (001) and (111) surfaces of NiMnSb, the appearance of surface
states destroys the half-metallicity.\cite{iosifSurf,iosif111} In Sections
\ref{sec3-1} and \ref{sec3-2} we present the FSKKR results for the
NiMnSb/InP(001) and NiMnSb/GaAs(001) contacts, respectively, and in
Section \ref{sec3-3} we give the valence band offsets calculated with
the FLAPW method for the NiMnSb/InP(001) interfaces and compare the
results obtained with the two different methods.

\begin{table}
\caption{
  FSKKR-calculated atomic spin moments given in $\mu_B$ for the
  interface between the MnSb-terminated (001)
  NiMnSb and the In or the P terminated InP. We do not present the
  spin moments at the vacant sites. Last columns are the moments for
  the MnSb-terminated NiMnSb(001) surface and the bulk NiMnSb.  ``I''
  denotes the interface layers and $\pm$ means one layer deeper in the
  half-metal or the semiconductor.}\label{table1}
  \begin{ruledtabular}
\begin{tabular}{rrrrr}
   & MnSb/In   & MnSb/P   &MnSb surf. & bulk     \\ \hline
I-3& Ni: 0.270 &Ni: 0.275 &Ni: 0.269  & Ni: 0.264\\
I-2& Mn: 3.704 &Mn: 3.734 &Mn: 3.674  & Mn: 3.705\\
I-2& Sb:-0.057 &Sb:-0.044 &Sb:-0.066  & Sb:-0.062\\
I-1& Ni: 0.289 &Ni: 0.316 &Ni: 0.223  & Ni: 0.264\\
I  & Mn: 3.405 &Mn: 3.718 &Mn: 4.018  & Mn: 3.705\\
I  & Sb:-0.037 &Sb:-0.045 &Sb:-0.096  & Sb:-0.062\\ \hline
I  & In:-0.053 &P : 0.015 & & \\
I+1& P :-0.022 &In:-0.013 & & \\
I+2& In:-0.017 &P :-0.011 & & \\
I+3& P :-0.012 &In: 0.002 & &
\end{tabular}
\end{ruledtabular}
\end{table}

\subsection{NiMnSb/InP contacts
\label{sec3-1} }

The first case which we will study are the interfaces between NiMnSb
and InP.  In Table \ref{table1} we have gathered the FSKKR spin
moments for the case of the MnSb/In and MnSb/P interfaces.  ``I''
stands for the interface layers, +1 means moving one layer deeper in
the semiconductor and $-1$ one layer deeper in the half-metallic spacer.
In the case of the MnSb terminated half-metallic film there is a
difference depending on the semiconductor termination. In the case of
the In termination the Mn spin moment decreases considerably and is
now 3.4 $\mu_B$ compared to the bulk value of 3.7 $\mu_B$.  For the P
terminated InP film the spin moment of Mn at the interface is very
close to the bulk value. In the case of the bulk NiMnSb the minority
gap is created by the hybridization between the $d$-orbitals of the Ni
and Mn atoms, but the Sb atom plays also a crucial role since it
provides states lower in energy than the $d$ bands which accommodate
electrons of the transition metal atoms.\cite{iosifHalf} Moreover Mn
and Ni atoms create a common majority band where there is a charge
transfer from the Mn atoms towards the Ni ones.  On the MnSb
terminated surface each Mn atom loses 2 out of its 4 nearest Ni atoms
and regains this charge which fills up mainly majority states. The Mn
spin moment at the surface is strongly enhanced reaching 4.0 $\mu_B$.
In the case of the interfaces, the final spin moment of the Mn atom at
the interface depends on the hybridization with the neighboring atoms
of the semiconductor. At an In interface, the Mn minority
$d$-states hybridize strongly with the In states and thus the Mn spin
moment is severely reduced and In shows a negative induced spin
moment. In the case of P, the situation is reversed and P has a
positive induced spin moment. The Mn-$d$ - P-$p$ hybridization is not
as strong as the Mn-$d$ - In-$p$ one and the Mn spin moment at the
interface is close to the bulk value.  We should also note that, if we
move deeper into the half-metallic film, the spin moments regain their
bulklike behavior while, if we move deeper in the semiconductor film,
the induced spin moments quickly vanish.

\begin{figure}
  \begin{center}
    \includegraphics[scale=0.33]{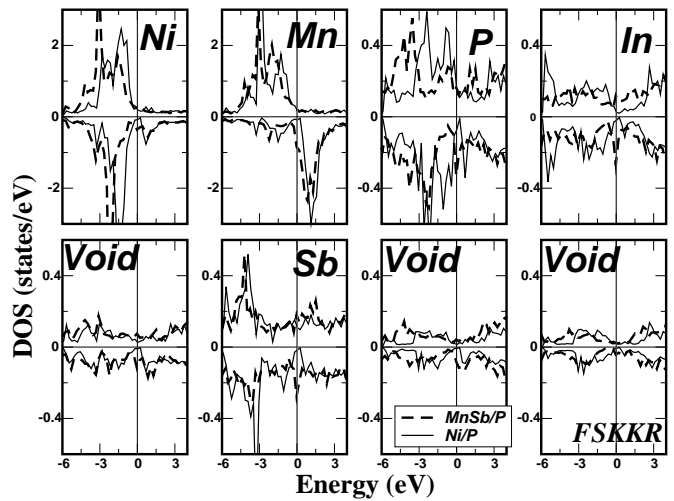}
  \end{center}
\caption{ \label{fig2}
  Atom- and spin-resolved DOS for the case of MnSb/P (dashed line) and
  Ni/P (solid line) contacts for the two interface layers and one
  layer deeper in the half-metal and the semiconductor. The zero of
  the energy is chosen to correspond to the Fermi level. Positive
  values of the DOS correspond to the majority spin and negative to
  the minority.}
\end{figure}

On a MnSb-terminated (001) surface the spin polarization at the Fermi
level, $E_F$, was found to be as high as 38\% (the spin-polarization
is defined with respect to the density of states $n(E)$:
$P=\frac{n^\uparrow (E_F)- n^\downarrow (E_F)} {n^\uparrow (E_F)+
  n^\downarrow (E_F)}$ where $\uparrow$ stands for the majority
electrons and $\downarrow$ for the minority electrons).
In Ref.~\onlinecite{Jenkins} two surface states at $E_F$ were
reported to destroy the half-metallicity, but still the population of
the majority electrons at the Fermi level was twice as large as
the one of the minority states.  Compared to this surface, for the
interfaces between MnSb-terminated NiMnSb and InP the situation is
completely different. The hybridization between the $d$-states of Mn
and $p$-states of Sb with the $p$-states of either the In or the P
atom at the interface is such that the net polarization at the
interface is almost zero. This is clearly seen in Fig.~\ref{fig2}
where we present with the dashed line the spin and atom resolved
density of states (DOS) of the atoms at the interface for a MnSb/P contact.
There is a minority interface state pinned at the Fermi level which
destroys the half-metallicity. In the Mn local DOS, this state overlaps
with the unoccupied minority Mn states and it is not easily
distinguished but its existence is obvious if one examines the Ni and
Sb DOS.  The situation is similar also for the MnSb/In contact not
shown here.

\begin{figure}
  \begin{center}
    \includegraphics[scale=0.35]{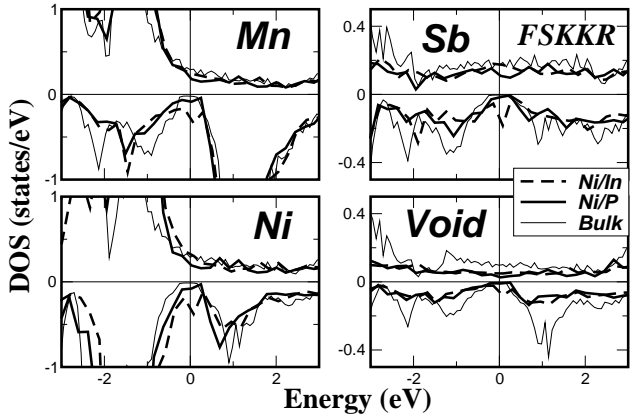}
  \end{center}
\caption{ \label{fig3}
  Bottom: Spin and atom-resolved DOS for the Ni and ``Void'' atoms at the
  interface with In (dashed line) or P (solid line). The top panels
  show the Mn and Sb DOS at the subinterface layer. The thin solid line
  indicates the results for bulk NiMnSb from Ref.~\onlinecite{iosifHalf}.}
\end{figure}

In the case of the Ni terminated NiMnSb films, DOS at $E_F$ is more
bulklike than the case of the MnSb films. Already Ni interface atom
has a spin moment of 0.29 $\mu_B$ in the case of an interface with In
and 0.36 $\mu_B$ for an interface with P compared to the bulk value of
0.26 $\mu_B$. In the bulk case Ni has 4 Mn and 4 Sb atoms as first
neighbors. On the Ni-terminated (001) surface the Ni atom
loses half of its first neighbors. But if an interface with
P is formed, the nickels two lost Sb neighbors are replaced by two
isovalent P atoms and -- with the exception of the Mn neighbors --
the situation is very similar to the bulk. Now the Sb $p$ bands at
lower energy are not destroyed  since P has a behavior similar to Sb
and still they accommodate three transition metal $d$ electrons.
Thus the only change
in the DOS comes from the missing two Mn neighboring atoms.  The DOS
in Fig.~\ref{fig2} for the Ni/P case is clearly very close to the bulk
case and in Fig.~\ref{fig3} we have gathered the DOS for the Ni and
the void at the interface and the Mn and Sb atoms at the subinterface
layer for both Ni/In (dashed line) and Ni/P (solid
line) contacts and we compare them with the bulk results from Ref.
\onlinecite{iosifHalf}.  In the case of the Ni/In interface there is
an interface state pinned at the Fermi level which completely suppresses
the spin polarization, $P$ (if we take into account the first two
interface layers is $P \approx 0$). In the case of the Ni/P interface the
intensity of these interface states is strongly reduced and now the
spin-polarization for the first two interface layers is 39\%, i.e.\
about 70\% of the electrons at the Fermi level are of majority
spin character.

\subsection{NiMnSb/GaAs contacts
\label{sec3-2} }

In the previous section it was shown that in the case of the Ni/P
interfaces the spin-polarization was as high as 39\%. In order to
investigate whether this is a general result for all semiconductors or
specifically for this interface we also performed calculations for the
case of the NiMnSb/GaAs(001) contacts using the same lattice parameter
as for the previous ones; thus the lattice constant of GaAs was
expanded by approximately 4\%.

\begin{figure}
  \begin{center}
    \includegraphics[scale=0.3]{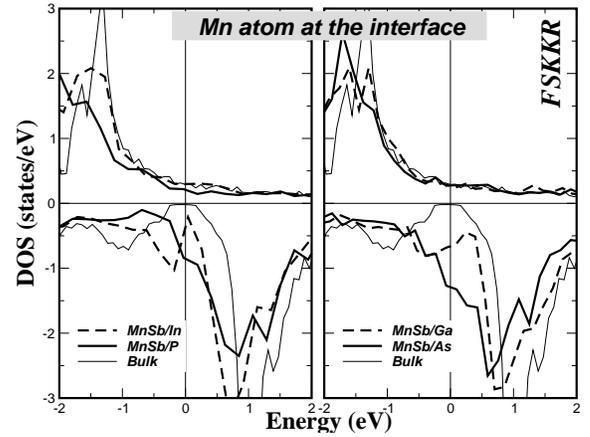} \vskip 0.5cm
    \includegraphics[scale=0.3]{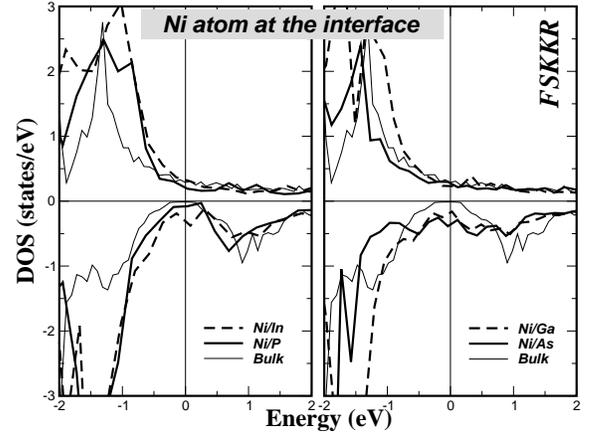}
  \end{center}
\caption{ \label{fig4}
  Left-top panel: atom- and spin-resolved DOS for the case of MnSb/In
  (dashed line) and MnSb/P (solid line) contacts for the Mn atom at
  the interface layer. Right-top panel: similar results for the
  MnSb/Ga and MnSb/As contacts. Bottom panels contain the results for
  the Ni-terminated half-metallic spacer. With the thin solid line
  the bulk results are indicated.}
\end{figure}

In the top panel of Fig.~\ref{fig4} the atom-resolved DOS for the Mn at
the interface layer is shown for the case of the
MnSb/semiconductor interfaces. The hybridization between the
$d$-orbitals of the Mn atom at the interface and the $p$-orbitals of
the $sp$ atoms of the semiconductor is larger in the case of the GaAs
than for the InP spacer. This leads to an about 0.1-0.2 $\mu_B$ smaller
Mn spin moments at the interface and the exchange
splitting between the occupied Mn majority and the unoccupied Mn
minority $d$-states is smaller.  Thus the large minority peak above
the Fermi level moves lower in energy and now strongly overlaps with
the occupied minority peak below the Fermi level increasing the
minority DOS at the Fermi level. In the meantime the smaller exchange
splitting causes the shift of the occupied Mn majority states towards
higher energies enhancing also the Mn majority DOS at the Fermi level.
The final spin-polarization at the Fermi level in the case of the
MnSb/Ga or As contacts is similar to the MnSb/In and MnSb/P ones and
these interfaces are not interesting for real applications.

The same effect occurring for the MnSb interface can also be seen
at the Ni interfaces, as shown in the bottom panels of Fig.~\ref{fig4}.
The stronger hybridization of the Ni atom with either the Ga or As atoms
at the interface with respect to the InP semiconductor provokes a
movement of the Ni unoccupied minority $d$ states towards lower
energies while the occupied majority ones are moving higher in energy.
If one looks in detail at the Ni/In and Ni/Ga contacts, one observes
that the minority peak at the Fermi level present in the Ni/In contact
is now smeared out in the case of the Ni/Ga contact due to the unoccupied
minority states which move lower in energy.  Similarly the unoccupied
Ni minority $d$-states have a larger bandwidth in the case of the
Ni/As contact than in the case of the Ni/P one inducing a high
minority Ni DOS at the Fermi level.  The high spin-polarization at the
Fermi level presented in the case of the Ni/P interfaces is completely
destroyed in the case of the Ni/As contact due to the larger
hybridization between the Ni-$d$ and As-$p$ orbitals with respect to
the hybridization between the Ni-$d$ and P-$p$ orbitals. Thus, the
properties of the interface depend also in a large extent on the
choice of the semiconductor.

\begin{figure}
  \begin{center}
    \includegraphics[scale=0.38]{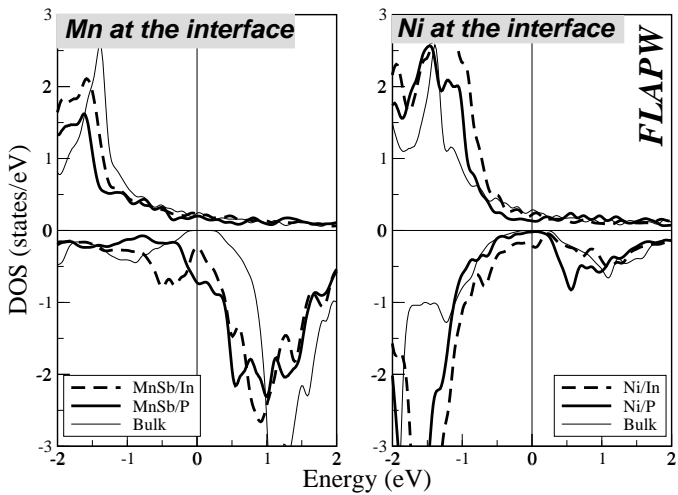}
  \end{center}
\caption{ \label{fig5}
  Left panel: Atom- and spin-resolved DOS for the
  case of MnSb/In(001) (dashed line) and MnSb/P(001) (solid line)
  contacts for the Mn atom at the interface layer as calculated
  with the FLAPW method. Right panel:
  Results for the Ni atom at the Ni/In and Ni/P interface for the
  Ni-terminated half-metallic spacers. The thin solid line
  indicates the local DOS of bulk NiMnSb.}
\end{figure}

\subsection{Band offsets and partial DOS for NiMnSb/InP contacts
\label{sec3-3} }

Employing the FLAPW method, we calculated the (minority states)
valence-band offset which is the energy difference between the
maximum of the valence band (VBM) of the semiconductor and the
maximum of the minority valence band of the Heusler alloy.  To
calculate it we referenced the binding energies the core states
in the interface calculation to their corresponding bulk values
as described in Ref.~\onlinecite{Freeman}.
We found  that the VBM of the semiconductor is 0.83 eV lower than
the one of the half-metal for the In/MnSb contact. For the other
interfaces the valence band offsets are: 0.69 eV for the In/Ni,
0.69 for the P/Ni and 0.80 eV for the P/MnSb contact.
In the bulk InP semiconductor the
experimental gap is 1.6 eV, thus the Fermi level, which is 0.07 eV
above the maximum of the minority NiMnSb valence band falls in the
middle of the semiconductor bulk bandgap. This is similar to what
is happening also in the case of the Co$_2$MnGe/GaAs (001)
interfaces~\cite{PicozziInter1} and these junctions
can be used to inject spin-polarized electrons in the semiconductor.

We can now also  compare the results obtained with the FLAPW
with the results from the FSKKR calculations. In the left panel of
Fig.~\ref{fig5} we present the DOS of the Mn atom at the MnSb/In and
MnSb/P interfaces together with the bulk FLAPW calculations while in
the right panel of the same figure we present the DOS for the Ni atom
at the Ni/P and Ni/In interfaces. We can directly compare these
results with the FSKKR results on the same systems shown in the
left-top and left-down panels of Fig.~\ref{fig4}. Except for very
small details both methods give a similar density of states.  In
the case of the MnSb/In interface the Fermi level falls within a local
minority minimum while for the MnSb/P interface, due to the smaller
exchange splitting Mn unoccupied minority states move lower in energy
crossing the Fermi level. More importantly both methods describe to
the same degree of accuracy the hybridization between the Ni
$d$ orbitals and the In or P $p$ states. For the Ni/In contact the Fermi
level is pinned within a minority Ni peak, the only difference being
that this peak is larger in the case of the FSKKR calculations. In the
case of the Ni/P interface the minority Ni DOS at the Fermi level is
very small as it was the case for the FSKKR results above.  Moreover
both methods yield similar spin magnetic moments at the interfaces
(and thus the spin moments calculated with the FLAPW method are not
presented here).

\begin{figure}
  \begin{center}
    \includegraphics[scale=0.32]{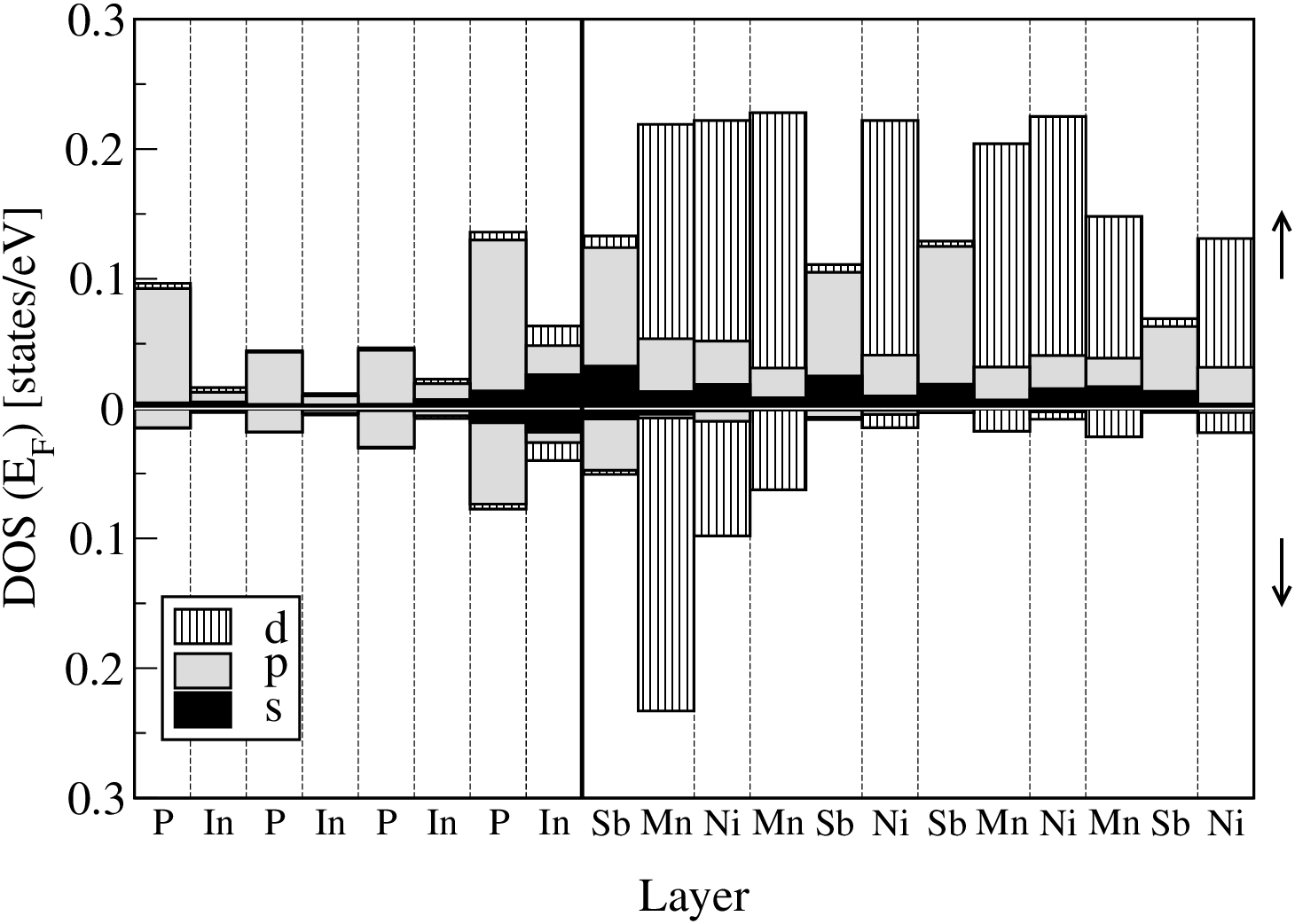} \vskip 0.5cm
    \includegraphics[scale=0.32]{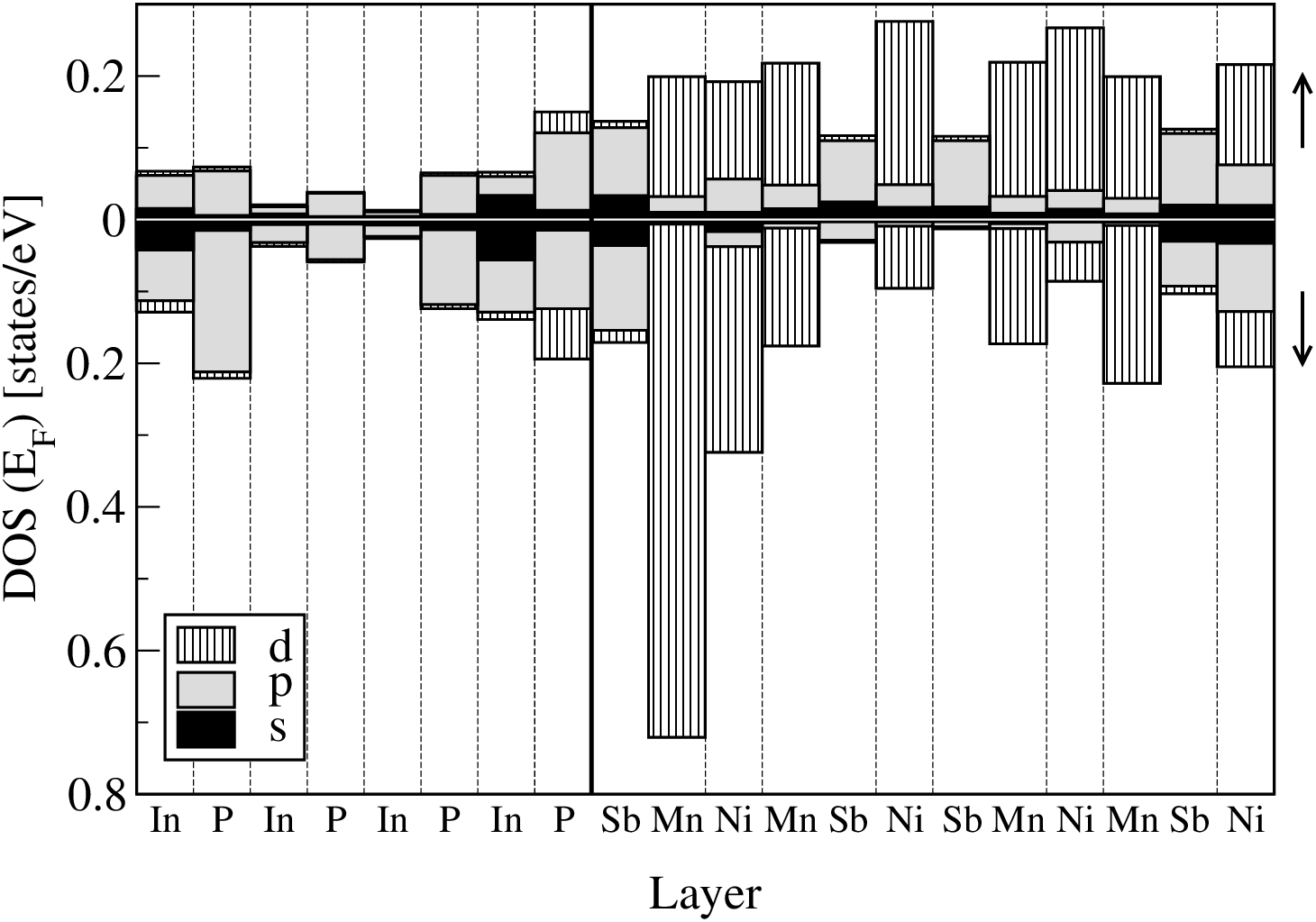}
  \end{center}
\caption{ \label{fig6}
  Layer-resolved DOS at the Fermi level for the (001) NiMnSb/InP
  contacts using the FLAPW method.
  Top: MnSb/In and Ni/P interfaces, bottom:  MnSb/P and Ni/In
  interfaces.
}
\end{figure}

To make our results more clear in Fig.~\ref{fig6} we have gathered the
layer-resolved partial DOS at the Fermi level for all the (001) interfaces
studied with the FLAPW method. As we already mentioned in
Section \ref{sec2}, we have used a slab made up from eight NiMnSb and
eight InP layers. Thus if one interface is MnSb/In (shown in the middle
of the top figure) then the other interface is Ni/P like and it consists
of the two layers shown at the edges of this figure (the slab is
periodically repeated along the axis perpendicular to the interface).
Similarly,  the bottom graph contains
the results for the MnSb/P and Ni/In interfaces. The layers at the
middle of the semiconductor spacer show a small DOS due to both the
induced states from the half-metal and bulk NiMnSb states which decay
slowly outside the half-metallic spacer and travel throughout the
semiconductor. It is clearly seen that none of the interfaces is in
reality half-metallic. For the MnSb/In interface the Mn atom at the
interface shows an almost zero net spin-polarization while the Mn atom
at the MnSb/P interface shows a quite large minority DOS as we have
already discussed.  In the case of the Ni/In interface shown in the
bottom panel of  Fig.~\ref{fig6},
the net spin-polarization is also almost zero as was the
case for the FSKKR results. The Ni/P interface shows a
spin-polarization $P$ around 40\% due to the high polarization of the
Ni atom at the interface which polarizes the P atom at the interface
also presenting a high majority DOS at the Fermi level.  This value is
almost identical to the 39\% calculated within the FSKKR method as
discussed in Section \ref{sec3-1}.

Although different types of slabs are used to describe the
interfaces and different approximations to the
exchange-correlation potential are employed both FSKKR and FLAPW
calculations lead to very similar results. The latter fact
confirms the results in Ref.~\onlinecite{calc5}, where it was
shown that for the same lattice constant both LDA and GGA
reproduce the same electronic and magnetic properties for the
Heusler alloys.

\section{N\lowercase{i}M\lowercase{n}S\lowercase{b}(111)/I\lowercase{n}P(111)
  interfaces
\label{sec4} }

In the last section we will discuss our FLAPW results for the
NiMnSb/InP (111) interfaces.  As mentioned in Section \ref{sec2},
for these calculations 16 layers of NiMnSb and 12 layers of InP
have been used. Along the [111] direction the semiconductor is
composed by pure alternating In and P layers and, thus, our
semiconducting spacer is ending in P on the one side and In on the
other side. In the case of the half-metallic alloy the structure
could be understood easier if we also assume that there is a
vacant site in the bulk structure (these ``voids'' have been
explicitly included in in the FSKKR calculations as described in
Section \ref{sec2} and Ref. \onlinecite{iosif111}).  For the Mn
termination, as we proceed from the interface deeper into the
half-metallic spacer, the succession of the layers can be either
Mn-Ni-Sb-Void-Mn-... or Mn-Void-Sb-Ni-Mn-... . We denote the two
different terminations here as Mn-Ni-Sb-Mn-... or Mn-Sb-Ni-Mn-...
. Similarly for the Sb terminated interface we can have either Mn
or Ni as subinterface layer and for the Ni termination we can have
either Sb or Mn at the subinterface layer. Since we have 16 layers
of the half-metallic alloy, we will have from both sides the same
layer at the interface, e.g.\ Mn, but with different subinterface
layers, e.g.\ Sb from one side and Ni from the other.

\begin{figure}
  \begin{center}
    \includegraphics[scale=0.32]{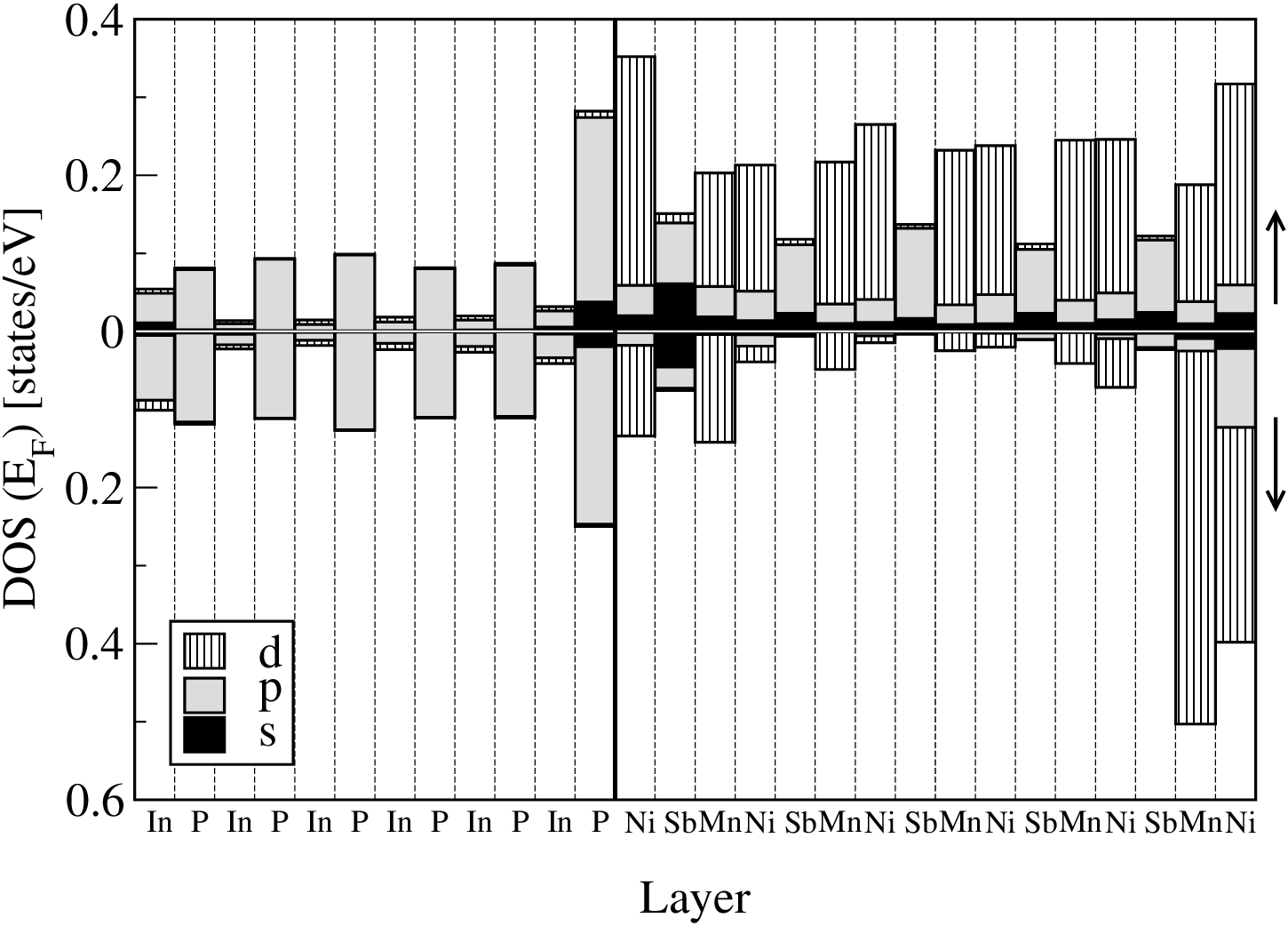} \vskip 0.5cm
    \includegraphics[scale=0.32]{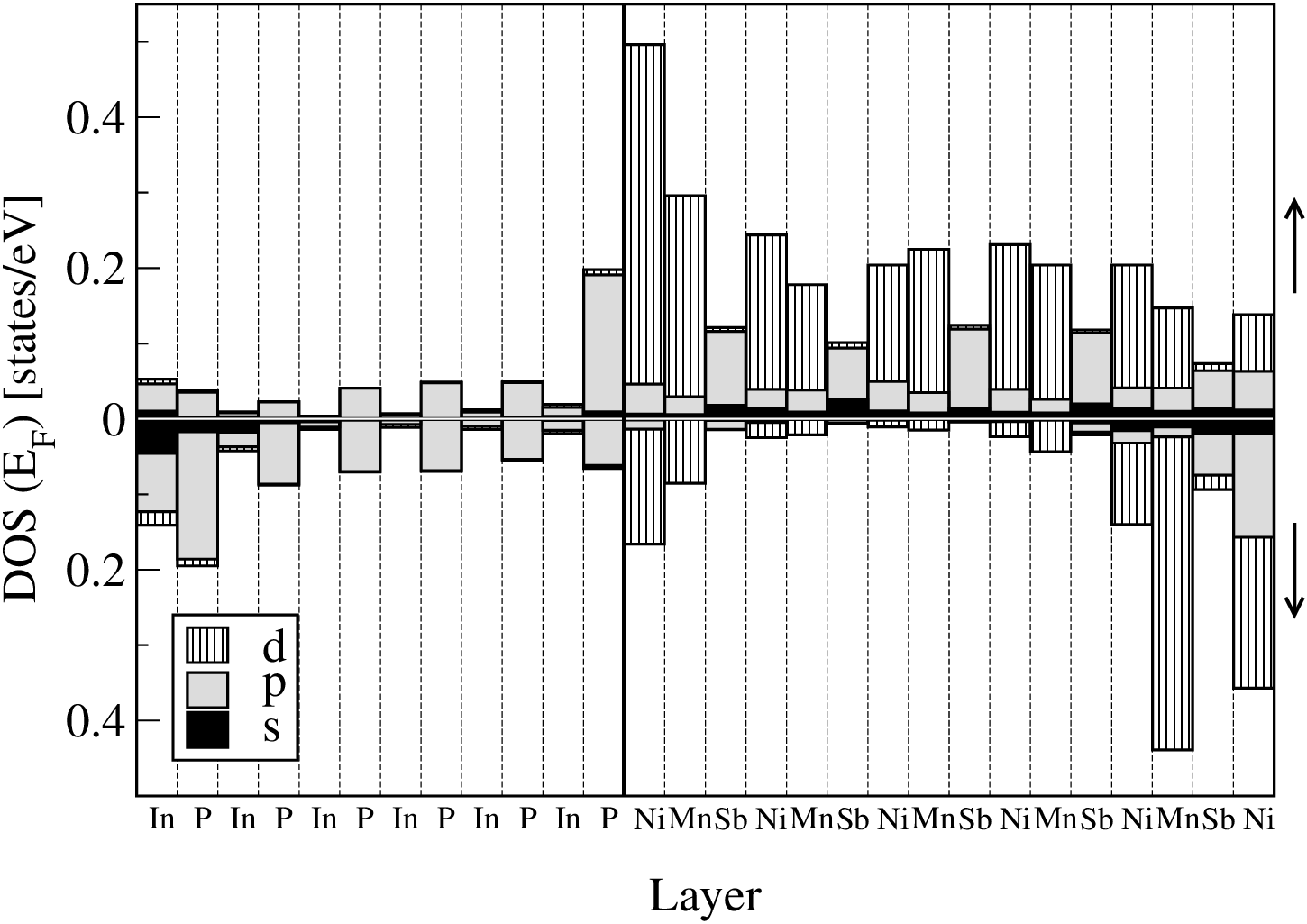}
  \end{center}
\caption{ \label{fig7}
  Layer-resolved DOS at the Fermi level for the Ni-terminated
  NiMnSb/InP(111) contacts calculated using the FLAPW method.
  In the middle of the figures a Ni/P interface is shown with
  Sb (top) or Mn (bottom) in the subinterface layer, while  at
  the borders of the figures the layers of a Ni/In interface
  can be seen with  Mn (top) or Sb  (bottom) in the
  subinterface layer.}
\end{figure}

In Fig.~\ref{fig7} the layer-resolved partial DOS for the
Ni-terminated interfaces are shown.  In the top panel are the
...-In-P/Ni-Sb-Mn-... and ...-P-In/Ni-Mn-Sb-... contacts and in the
bottom panel the ...-In-P/Ni-Mn-Sb-... and ...-P-In/Ni-Sb-Mn-... ones.
As it was shown in Ref.~\onlinecite{iosif111}, in the case of the Ni
and Mn terminated (111) surfaces there are strong surface states
pinned at the Fermi level which also penetrate deeply into the subsurface
layers. These surface states are present also in the case of the
interfaces studied here, although their intensity decreases slightly
due to the hybridization with the $sp$ atoms of the semiconductor. In
all cases the net spin-polarization of the Ni atom at the interface is
very small with the exception of the ...-In-P/Ni-Mn-Sb-... interface
(middle of the bottom panel). For this case the simultaneous presence
of the P atom from the one side and of the Mn atoms at the
subinterface layer create an atomic-like environment for Ni similarly
to what happened in the case of the Ni/P(001) contact and the
spin-polarization, taking into account the two semiconductor layers at
the interface and three first NiMnSb layers, is as high as $\sim$53\%
and thus more than 76\% of the electrons at the Fermi level are of
majority character. In the case of the Mn-terminated NiMnSb-films (not
shown here) the interface states are even stronger than for the
Ni-terminated spacers and the spin-polarization at the interface
vanishes.

\begin{figure}
  \begin{center}
    \includegraphics[scale=0.32]{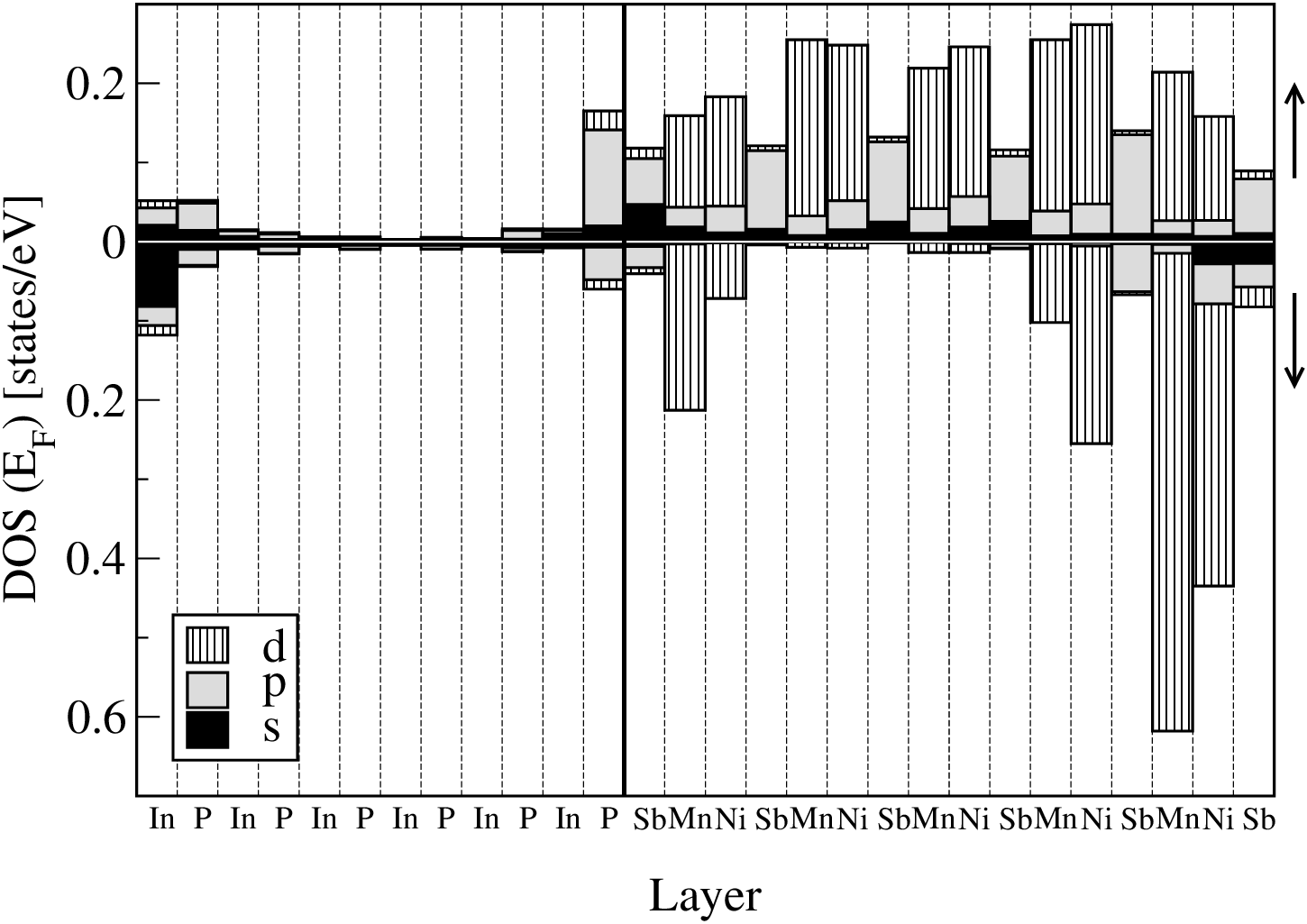} \vskip 0.5cm
    \includegraphics[scale=0.32]{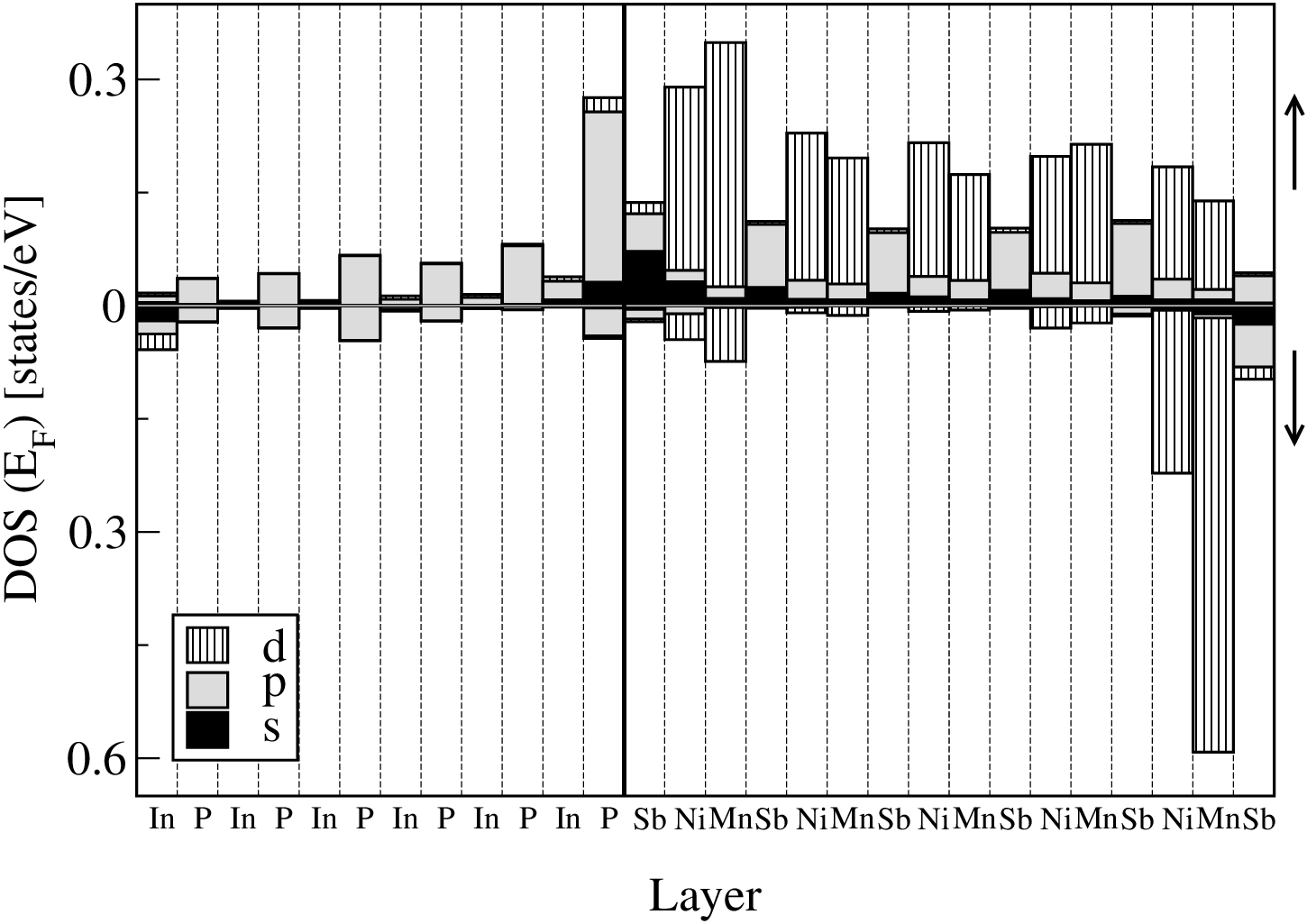}
  \end{center}
\caption{ \label{fig8}
  Layer-resolved DOS at the Fermi level for the Sb-terminated
  NiMnSb/InP(111) contacts calculated using the FLAPW method.
  In the middle of the figures a Sb/P interface is shown with
  Mn (top) or Ni (bottom) in the subinterface layer, while  at
  the borders of the figures the layers of a Sn/In interface
  can be seen with  Ni (top) or Mn  (bottom) in the
  subinterface layer.}
\end{figure}

In the last part of our study we will concentrate on the Sb-terminated
(111) interfaces In their paper Wijs and de Groot predicted
that the interfaces between the Sb-terminated NiMnSb(111) film and a
S-terminated CdS(111) film should keep the half-metallicity or at least
show an almost 100\% spin-polarization at the Fermi
level.\cite{groot2} Thus it is of particular interest to study the
interfaces between the Sb atom and  P, although P has one
electron less than S. Firstly, we should note that contrary to the Mn
and Ni terminated, in the case of the Sb- terminated NiMnSb(111)
surfaces the interface state was not pined exactly at the Fermi level
but slightly below it and the spin-polarization in the case of the Sb
surfaces was still high.\cite{iosif111} In the case of the interfaces
between In and Sb half-metallicity is completely destroyed and the
spin-polarization is even negative; there are more minority-spin
electrons at the Fermi level than majority ones as can be seen from
the DOS at the boundaries of the pictures in Fig.~\ref{fig8}.

In Fig.~\ref{fig8} we also show the two different P/Sb-terminated
interfaces: In the top panel the one with Mn as subinterface layer
is not of particular interest since the Mn atom shows a
practically zero net spin-polarization decreasing considerably the
overall spin-polarization at the interface. On the other hand,
when the subinterface layer is Ni as in the middle of the bottom
panel, all atoms at the interface show a very high majority DOS at
the Fermi level and the resulting spin-polarization, $P$, is
$\sim$74\% and thus $\sim$86\% of the electrons at the Fermi level
are of majority character.  We should also mention that, although
the induced majority DOS at the Fermi level for the P atom at the
interface seems very large (it is of the same order of magnitude
with the Ni one), when we move away from the Fermi level it
becomes very small compared to the majority DOS of the
transition-metal atoms.

\begin{figure}
  \begin{center}
    \includegraphics[scale=0.45]{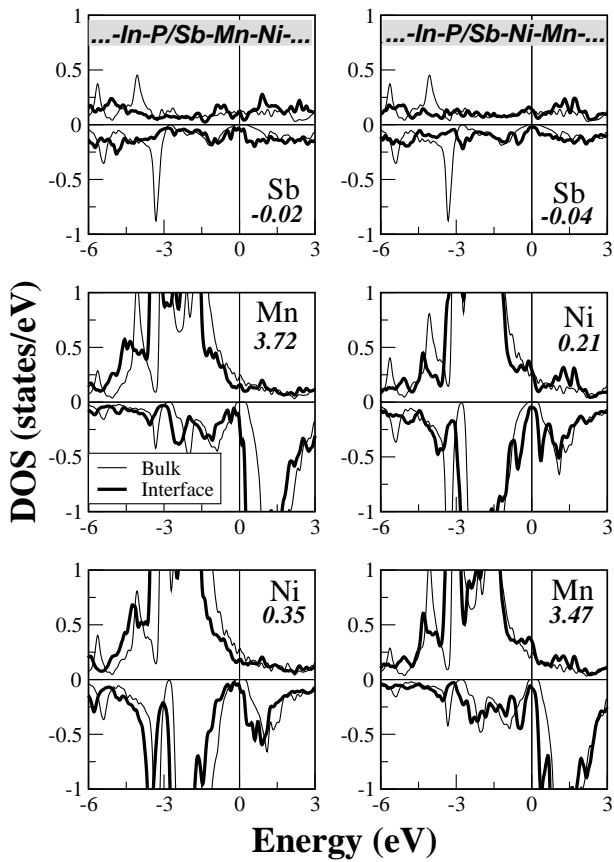}
  \end{center}
\caption{ \label{fig9}
  Atom- and spin-resolved DOS for the ...-In-P/Sb-Mn-Ni-... interface
  (left panels) and the ...-In-P/Sb-Ni-Mn-... interface (right panels)
  calculated with the FLAPW method. The values in the figures
  are the spin moments of the atoms at the interface in $\mu_B$. The
  thin solid line indicates the bulk results.}
\end{figure}

The main question needed still to be answered is why the two different
P/Sb interfaces show such large differences. It is mainly the Mn
atom whose spin-polarization at the Fermi level is very different
depending of its distance from the interface.  To answer this question
in Fig.~\ref{fig9} we have gathered the layer-resolved DOS for the two
different P/Sb interfaces and in this figure we also included the
atomic spin moments.  The Sb spin moments are $-0.02 \mu_B$ for the
...-In-P/Sb-Mn-Ni-... interface and $-0.04 \mu_B$ for the
...-In-P/Sb-Ni-Mn-... interface, in both cases this is smaller than
the bulk value of $-0.06 \mu_B$.
The Mn spin moment for the ...-In-P/Sb-Mn-Ni-... case
is 3.72 $\mu_B$, close to the bulk value of 3.70 $\mu_B$, considerably
larger than the Mn moment of 3.47 $\mu_B$ for the
...-In-P/Sb-Ni-Mn-...  case. One would expect that in the first case
the exchange splitting should be larger and the unoccupied minority
states would be higher in energy but, as can be seen in Fig.~\ref{fig9},
the contrary effect occurs. In the second case the Mn is deeper
in the interface and its environment is more bulklike and the minority
states are pinned at their position and thus the the Fermi level falls
within a minority local minimum resulting in a very high
spin-polarization.
At the  ...-In-P/Sb-Mn-Ni-... contact  the Mn atom is closer to the
interface. Here, the larger hybridization
of the Mn minority states (not only with the $p$-orbitals of Sb but also
with the ones of $P$, since the last ones are closer now) obliges the
minority states to move slightly lower in energy. Thus, the Fermi
level does not fall in the local minimum but shifts into  the peak of the
unoccupied minority states and the net spin-polarization vanishes.
The Ni states are strongly polarized by the Mn ones and also in the
case of the Ni atom which is deeper than the Mn one, the Fermi level
does not fall anymore within the local minimum.

Similarly to the (001) interfaces in Section \ref{sec3-3}, we also
calculated the band-offset in the case of the (111) interfaces. The
band-offset ranges from 0.36 eV in the case of the
...-In-P/Mn-Sb-Ni-...  contact up to $\approx$1 eV for the
...-In-P/Sb-Ni-Mn-... configuration. Thus the conclusions of Section
\ref{sec3-3} are valid also for these interfaces.

\section{Summary and conclusions
\label{sec5} }

In the first part of our study we investigated the electronic and
magnetic properties of the (001) interfaces between the half-metal
NiMnSb and the binary semiconductors InP and GaAs using two different
full-potential \textit{ab-initio} techniques. Both methods gave similar
results in the case of the NiMnSb/InP(001) contacts. In all cases the
(001) interfaces lose the half-metallicity but in the case of the Ni/P
contact the Ni has a bulk like behavior since the P atoms substitute
the cut-off Sb isovalent neighbors and 70\% of the electrons are of
majority-spin character at the Fermi level. But in the case of the
Ni/As interface the large hybridization at the interface suppresses this
high spin-polarization. MnSb-terminated interfaces, on the other hand,
present very intense interface states which penetrate also into the
deeper layers of the NiMnSb film.

In the second part of our study we investigated  all the
possible (111) interfaces between NiMnSb and InP. In all cases
interfaces states destroy the half-metallicity but in two cases the
interface presents high spin-polarization. Firstly, when the contact is
the ...-In-P/Ni-Mn-Sb-..., the Ni atom at the interface has a bulklike
environment and the spin-polarization at the Fermi level is as high as
53\%. In the case of the ...-In-P/Sb-Ni-Mn-... contact the
spin-polarization is even higher reaching a value of 74\%.

Although half-metallicity at the interfaces is in general lost, there
are few contacts in which a high spin-polarization remains, that makes them
attractive for realistic applications.  Interface states are important
because the interaction with defects makes them conducting and
lowers the efficiency of devices based on spin-injection. Thus,
building up interfaces with the highest spin-polarization possible
like the ones proposed here is a perquisite but not a guarantee to
achieve highly efficient spin-injection.

\acknowledgements

This work was financed in part by the BMBF under auspices of the
Deutsches Elektronen-Synchrotron DESY under contract no. 05 KS1MPC/4.

\end{document}